\documentclass{elsart}
\usepackage{graphicx}
\usepackage{amssymb}
\journal{Physics Letters A}
\begin{document}
\begin{frontmatter}
\title{Kraus decomposition for chaotic environments}

\author{Murat \c{C}etinba\c{s}\corauthref{cor}},
\corauth[cor]{Corresponding author.} 
\ead{cetinbas@sfu.ca}
\author{Joshua Wilkie}
\ead{wilkie@sfu.ca}
\address{Department of Chemistry, Simon Fraser University, Burnaby, British Columbia V5A 1S6, Canada}

\begin{abstract}
We consider a system interacting with a chaotic thermodynamic bath. We derive an explicit and exact Kraus operator sum representation (OSR) 
for the open system reduced density. The OSR preserves the Hermiticity, complete positivity and norm. We show that it is 
useful as a numerical tool by testing it against exact results for a qubit interacting with an isolated flawed 
quantum computer. We also discuss some interesting  qualitative aspects of the OSR.
\end{abstract}

\begin{keyword} Kraus decomposition \sep decoherence \sep coherent shift \sep chaotic baths
\PACS 03.65.$-$w \sep 05.30.$-$d \sep 03.67.Lx \sep 03.65.Yz \sep 05.45.Mt 
\end{keyword}
\end{frontmatter}

\section{Introduction}
Efforts to design scalable quantum computers (QCs) \cite{QC} and other new quantum technologies \cite{SB,Rat} are rekindling interest in the phenomenology of open quantum system dynamics.
While effects of atom-radiation field interactions \cite{Gard,Obath} have been extensively studied, the
consequences of interaction with a complex environment are just beginning to be explored. Solid and condensed phase environments bequeath a diversity of controllable interactions for scalable implementation of new quantum technologies, and thus understanding the open system dynamics induced by such complex environments is of particular interest. Recent studies have reported a number of interesting effects such
as suppression of decoherence with increasing bath chaos \cite{Tess,Chaobath,CW1,CW2} and large coherent 
shifting \cite{CW1,CW2}.

Manifestation of chaos \cite{Chaos,Chaos2}, or at least non-integrability, is a defining and unavoidable feature of complex environments.
Evidence suggests that condensed phase dynamics is generally chaotic: Dynamics 
of a colloidal particle in water \cite{Gasp}, and both vibrational dynamics \cite{Miya} of Si and its electronic 
structure \cite{Mucc} have been shown to be chaotic. Anharmonic corrections are known to be important in the study 
of phonons and essential for an understanding of heat transport \cite{Wol}. Thus, the chaotic bath model provides a better 
representation for the condensed or solid state environment than standard uncoupled oscillator bath models.

Chaotic bath models are substantially more difficult to simulate than the radiation field. At high temperatures 
random matrix theories \cite{Bulg}, or classical trajectory simulations such as the Wigner method \cite{GB}, 
can be used. Redfield theory \cite{Red,Opp} and its generalizations \cite{Coal,Neu} can also be employed. 
But at the very low temperatures relevant to quantum computing technologies the options are more limited. There is a recently
developed approximate master equation \cite{CW1,CW2,SRA1,SRA2,SRA3} which appears to be accurate in this low temperature regime. However,
as with any approximate theory its predictions need to be otherwise confirmed.

In this manuscript we derive a simple formula for the reduced density which captures all the important features of 
open system dynamics induced by chaotic environments in the low temperature regime. The equation is of the Kraus 
Operator Sum Representation (OSR) \cite{Kraus} form which automatically satisfies all the required conservation 
laws for the reduced density, i.e. Hermiticity, complete positivity and norm conservation.
Our Kraus OSR\cite{Kraus} provides a formally exact equation for the reduced density in terms of explicit representations
of the Kraus operators. Aside from chaos, the only required property of the bath Hamiltonian is that it be of
large thermodynamic dimension.

As a first test, we compare the predictions of the chaotic Kraus OSR\cite{Kraus} to exact calculations for a model of a 
single qubit interacting with an isolated QC with static internal flaws. While our derivation assumes
a bath of thermodynamic dimensions, our numerical test shows that very accurate results can be obtained even for a chaotic bath of just ten 
qubits.

Organization of this manuscript is as follows. A derivation of the Kraus OSR\cite{Kraus},in three steps, is given in Section 2. In section 3 we discuss our chaotic spin-bath model. In Section 4 we test the OSR against exact numerical results. In section 5 we discuss our results.

\section{Derivation of chaotic Kraus OSR}

Consider the following total Hamiltonian for the bipartite closed system
\begin{equation}
\label{Ham}
\hat{H}=\hat{H}_{S}+\hat{S}\hat{B}+\hat{H}_{B}
\end{equation}
where $\hat{H}_{S}$ is the system Hamiltonian and $\hat{H}_{B}$ is the bath Hamiltonian. $\hat{S}\hat{B}$ is the interaction Hamiltonian where $\hat{S}$ and $\hat{B}$ are the system and bath coupling operators, respectively.

Our derivation consists of three parts. First we review the construction of the OSR assuming a complete
bath eigenbasis (i.e $\hat{H}_{B}| j\rangle=E_{j}| j \rangle$ and $\sum_j|j\rangle\langle j|=\hat{I}_B$). Next we show that the off-diagonal ($j\ne k$)
matrix elements of bath coupling operator 
\begin{equation}
B_{j,k}=\langle j|\hat{B}|k\rangle\label{Bjk}
\end{equation}
vanish for a chaotic bath in the thermodynamic limit. Finally, we 
use this fact to show that the resulting Kraus OSR takes a particularly simple and potentially useful form.

\subsection{General Kraus OSR for complete bath eigenbasis}

We consider an uncorrelated initial state for the whole system
\begin{equation}
\hat{\rho}(0)=\hat{\rho}_{S}(0)\otimes \hat{\rho}_{B}(0)
\end{equation} 
where $\hat{\rho_{S}}(0)$ is an arbitrary initial state of the system and $\hat{\rho}_{B}(0)$ is an initial bath state of canonical form, i.e.
\begin{equation}
\hat{\rho}_{B}(0)=\sum_j \frac{ e^{-E_{j}/kT }} {Q} | j\rangle \langle j |. 
\end{equation}
The subsequent states will be given by the exact solutions 
\begin{equation}
\hat{\rho}(t)=\hat{U}(t)\hat{\rho}(0)\hat{U}^{\dagger}(t)
\end{equation}
 of the Liouville-von Neumann equation 
 \begin{equation}
 d \hat{\rho}(t) / dt=-i[\hat{H},\hat{\rho}(t)]. 
\end{equation}
Here $\hat{U}(t)=\exp{\{ -(i/\hbar) \hat{H} t \}}$. 
The exact reduced density of the system is then formally obtained by tracing over the bath degrees of freedom, i.e,
\begin{equation}
\hat{\rho}_{S}(t)={\rm Tr}_{B}\{ \hat{U}(t)\hat{\rho}(0)\hat{U}^{\dagger}(t) \}.
\end{equation}
Since the bath eigenbasis is complete, and the initial canonical bath density is already diagonal in eigenstates of $\hat{H}_{B}$, performing the partial trace in the same basis yields
\begin{equation}
\hat{\rho}_{S}(t) = \sum_{j,k} \langle j| \hat{U}(t) \left( \hat{\rho}_{S}(0) \otimes \frac{ e^{-\beta E_{k} }} {Q} | k\rangle \langle k | \right) \hat{U}^{\dagger}(t) | j \rangle
\end{equation}
which in turn can be written in more compact Kraus OSR form
\begin{equation}\label{OSR}
\hat{\rho}_{S}(t)=\sum_{j,k} \hat{\cal K}_{j,k}(t) \hat{\rho}_{S}(0) \hat{\cal K}_{j,k}^{\dagger}(t)
\end{equation}
where the Kraus operators are defined by
\begin{equation}\label{Kraus}
\hat{\cal K}_{j,k}(t)=\sqrt{p_{k}} \langle j| \hat{U}(t) | k\rangle.
\end{equation} 
Here $p_{k}=\exp{ \{ -\beta E_{k} \} }/Q$ are the initial populations of the bath eigenstates. Note that
$\sum_{j,k} \hat{\cal K}_{j,k}(t)\hat{\cal K}_{j,k}^{\dagger}(t)=\hat{I}_{S}$.   

This form of the Kraus OSR is of formal interest only. Although it is exact and so satisfies all 
the required conservation laws (i.e. the Hermiticity, complete positivity, and norm) for the subsystem density
it is not practical for computational purposes. First, the explicit forms of the Kraus operators (\ref{Kraus}) are not
known in general, because the Kraus OSR (\ref{Kraus}) is not unique. And secondly, the double sum in (\ref{OSR}) would make use of the OSR impractical even if simple forms
for (\ref{Kraus}) were known. We will now show that in the limit of a chaotic bath of thermodynamic dimension, explicit
forms for (\ref{Kraus}) can be supplied and that the double sum reduces to a unique single sum. 

\subsection{Chaotic bath of thermodynamic dimension}
Using the bath eigenvalues and eigenstates, the total Hamiltonian (\ref{Ham}) can alternatively be written as
\begin{equation}
\label{Ham2}
\hat{H}=\hat{H}_{S}+\hat{S}\sum_{j,k}B_{j,k} | j \rangle \langle k |+\sum_{j}E_{j} | j
\rangle \langle j |
\end{equation}
where $B_{j,k}$ are the bath coupling operator matrix elements defined above in Eq. (\ref{Bjk}), and $E_j$ and $|j\rangle$
are again the bath eigenenergies and eigenstates.

Decoherence and open system dynamical effects are mediated by the matrix elements $B_{j,k}$ and by operators $\hat{S}$ and $| j \rangle \langle k |$. We will
focus on the properties of $B_{j,k}$. Now we assume that $\hat{H}_{B}$ is chaotic (both quantally and classically) and that the Wigner function\cite{DS}
of $\hat{B}$ has a classical limit. Coordinate and momentum bath coupling operators, for example, have Wigner functions with
classical limits. Let $N$ denote the number of bath degrees of freedom. We will eventually send $N\rightarrow\infty$ to attain the thermodynamic limit.

Quantum chaos has a number of important consequences for matrix elements and spectra\cite{Chaos,Chaos2}. We will now show that off-diagonal
matrix elements vanish for thermodynamic baths. We begin with an exact formula for the off-diagonal matrix elements 
of an arbitrary quantum system:
\begin{eqnarray}
&&|B_{j,k}|^2=(\frac{h}{I_{j,k}})^N\int d{\bf x}~d{\bf x}_0 ~B({\bf x})^*B({\bf x}_0)\nonumber \\
&&\int d {\bf y} ~e^{2\pi i ({\bf x}-{\bf x}_0)J{\bf y}} ~W_j({\bf x}+h {\bf y}/2)W_k({\bf x}_0-h {\bf y}/2).\label{Bij2}
\end{eqnarray}
This is obtained by combining the exact equations (50) and (119) of \cite{SemiC1}.
Here $B({\bf x})$ is the Wigner function\cite{DS} of operator $\hat{B}$ and 
\[
W_j({\bf x})=C_j\int d{\bf v} ~e^{2\pi i {\bf p}{\bf v}}\langle {\bf q}-h{\bf v}/2|j\rangle\langle j|{\bf q}+h{\bf v}/2\rangle
\]
is a scaled Wigner function for eigenstate $j$. It is well known\cite{Berry} that $W_j({\bf x})$ has a well defined classical limit
if the classical Hamiltonian $H_B({\bf x})$ is chaotic. [The Wigner function of $\hat{H}_B$ may differ from the classical Hamiltonian by small corrections which vanish with Planck's constant, but we ignore this distinction in 
our discussion.] We employ
a notation where ${\bf x}=({\bf p},{\bf q})$, where ${\bf p}$ are the momenta and ${\bf q}$ the coordinates. $J$ is the
$2N\times 2N$ dimensional symplectic matrix. Here $C_j=\sqrt{\int d{\bf x} ~\delta (1-H_B({\bf x})/E_j)}$ and
$I_{j,k}=(C_jC_k)^{1/N}$ has the dimensions of action. 

Now, note that $\int d{\bf x}~ \delta (1-H_B({\bf x})/E_j)\propto E_je^{S_j/k_B}$ where $S_j$ is the microcanonical entropy and 
$k_B$ is the Boltzmann constant. Now since entropy is extensive, $S_j$ scales linearly with $N$ for sufficiently large $N$. Consequently, by l'H\^{o}pital's rule, 
it follows that $C_j$ scales exponentially with $N$. Finally, it follows that $I_{j,k}$ is constant, or at most grows weakly with $N$ since the energies $E_j$ also grow with $N$.

The Wigner functions $W_j({\bf x})$ and $B({\bf x})$ and the associated integrals in (\ref{Bij2}) have well defined classical\cite{Berry} and therefore thermodynamic limits. In the classical limit, for example, these terms reduce to $E_k\delta(E_j-E_k)\overline{B^2}$ which increases slowly with $N$. (Here $\overline{B^2}$ is the classical canonical average of the squared classical limit of the Wigner function for $\hat{B}$. Note that $\hat{B}\propto \sqrt{N}$ and so $\overline{B^2}$ would normally scale linearly with $N$. $E_k\delta(E_j-E_k)$ is only weakly dependent on $N$. Note also that $\delta(E_j-E_k)$ should be understood as a distribution with a small but nonzero width.) Thus, by l'H\^{o}pital's rule, the magnitude of $|B_{j,k}|^2$ in the thermodynamic limit is determined by the factor $(\frac{h}{I_{j,k}})^N$. Consequently, if $h<I_{j,k}$, which will always be true for sufficiently large $N$, then $|B_{j,k}|^2\rightarrow 0$ exponentially fast as $N\rightarrow \infty$. Note that the requirement is $h<I_{j,k}$, not the stronger semiclassical $h< <I_{j,k}$.

Hence, the off-diagonal matrix elements of $\hat{B}$ vanish for a chaotic bath of thermodynamic dimension.
They also decay exponentially fast with increasing $N$, and so they may be neglected even for quite small baths.

In the semi-classical limit, the scaling, $|B_{j,k}|^2 \propto h^{N}$ has long been known in the quantum chaos literature\cite{SemiC1,SemiC2}, but its consequences in the thermodynamic limit, and for the Kraus OSR in particular, have not been 
appreciated. As we have shown, the semi-classical limit $h \rightarrow 0$ is not a necessary condition for the vanishing of the off-diagonals, $|B_{j,k}|^2 \rightarrow 0$, for a chaotic bath of thermodynamic dimension. 

It is also worth noting that similar formulas arise in other contexts from completely different arguments\cite{HPB}. 

\subsection{Chaotic Kraus decomposition}
Now, we will show that the vanishing of the off-diagonal matrix elements of the bath coupling operator has 
important consequences for the Kraus OSR (i.e.  Eq. (\ref{OSR})). To begin with, Hamiltonian (\ref{Ham2}) 
reduces the simpler form
\begin{equation}
\label{Ham3}
\hat{H}=\hat{H}_{S}+\sum_{k}(\hat{S}B_{k,k}+E_{k}) | k \rangle \langle k |.
\end{equation}
For integer powers of $l=0,..,\infty$ then, it is true that
\begin{equation}\label{Ham4}
\langle j| \hat{H}^{l} | k\rangle = \left( \hat{H}_{S}+ \hat{S}B_{k,k}+E_{k} \right)^{l} \delta_{j,k}
\end{equation}
for all $j$ and $k$. Using Eq. (\ref{Ham4}) for 
\begin{equation}
e^{-\frac{i}{\hbar}\hat{H}t}=\sum_{l=0}^{\infty}\frac{ (-i\hat{H}t)^{l}}{\hbar^{l}l!},
\end{equation}
one can readily show that the Kraus operators (\ref{Kraus}) take the explicit form 
\begin{equation}
\hat{{\cal K}}_{j,k}(t)=\sqrt{p_{k}}e^{ -\frac{i}{\hbar} (\hat{H}_{S}+\hat{S}B_{k,k} + E_{k}) t } \delta_{j,k} .
\end{equation}
Finally, substituting this back into (\ref{OSR}) gives the desired form of the chaotic Kraus decomposition 
\begin{equation} 
\hat{\rho}_{S} (t)=  \sum_{k} \frac{e^{ -\beta E_{k} }}{Q}  e^{-\frac{i}{\hbar}(\hat{H}_{S}+\hat{S}B_{k,k})t}\hat{\rho}_{S}(0) e^{\frac{i}{\hbar}(\hat{H}_{S}+\hat{S}B_{k,k})t} .\label{CTK}
\end{equation}

Equation (\ref{CTK}) is an exact equation for the reduced density of an arbitrary quantum system interacting with
a chaotic bath of thermodynamic dimension. While the arguments of the previous section were cast in terms of a phase space representation, 
similar arguments could be devised for spin baths using generalized coherent state representations\cite{SRA2}. In the
next section we actually employ a spin bath to test the formula (\ref{CTK}) numerically. We will show that (\ref{CTK})
is very accurate even for a bath of just ten qubits.

\section{Test model}

In our recent investigation \cite{CW1} we showed that internal decoherence, dissipation and coherent shifting, caused by static one and two-body internal imperfections, are potential error sources for flawed QCs\cite{GS1}. We also showed that a single detector qubit can be configured to probe internal bath dynamics \cite{CW2} and thus one can obtain valuable information about the bath self-interactions. Here we extend our previous exact numerical results to a longer time limit and test the chaotic Kraus decomposition against these exact results. In what follows we briefly review our flawed QC model and our exact numerical approach. Details of this study are given in \cite{CW2}.

\subsection{Exact numerical approach}

Our chaotic spin-bath model represents a flawed QC core\cite{GS1}. The total Hamiltonian takes the form
\begin{eqnarray}
\label{QCHam}
\hat{H}&=&-\frac{1}{2}{B}_{0}^{z} \hat{\sigma}_{z}^{(0)}
+ {\sigma}_{x}^{(0)} \sum_{i=1}^{N}\lambda_{i} \hat{\sigma}_{x}^{(i)} \\
&-&\frac{1}{2}\sum_{i=1}^{N} \left( B_{i}^{x}\hat{\sigma}_{x}^{(i)} 
+ B_{i}^{z} \hat{\sigma}_{z}^{(i)} \right)
+ \sum_{i=1}^{N-1}\sum_{j=i+1}^{N} J_{x}^{i,j} \hat{\sigma}_{x}^{(i)}\hat{\sigma}_{x}^{(j)},
\nonumber
\end{eqnarray}
\noindent
where the first term represents a central qubit (i.e. the detector qubit). The second term is the system-bath coupling operator, and the last two terms constitute the chaotic bath Hamiltonian which represents the rest of the flawed QC.  

Our numerical simulations are based on the experimentally relevant parameter regime for a Josephson charge-qubit QC proposal \cite{Nori}. We set the detector qubit energy to $B_{0}^{z}=1\  \epsilon$ in units of $\epsilon=200$ mK. We assume that all other qubits other than the detector qubit have one-body flaws, i.e. $B_i^z \in [ B_0^z-\delta/2,B_0^z+\delta/2 ]$ and similarly $B_i^x\in [B_0^z-\delta/2,B_0^z+\delta/2 ]$ where the detuning is set to $\delta=0.4\ \epsilon$. Two-body flaws, i.e. residual system-bath couplings, and chaos generating intra-bath interactions are modeled via $\lambda_{i} \in \ [-\lambda,\: \lambda]$ and  $J_{x}^{i,j} \in \ [-J_{x},\: J_{x}]$, respectively. We kept the bath temperature constant at $kT=0.25\ \epsilon$ in all our simulations. Keeping the value of the system-bath coupling constant at $\lambda=0.05\ \epsilon$ we considered three different values of intra-bath couplings; $J_{x}=0.50, \: 1.00, \:2.00$ in units of $\epsilon$. By $J_x=2\epsilon$ the
bath is fully chaotic. 
 
Exploiting the experimentally relevant low temperature regime \cite{Makhlin} for the charge-qubit QC proposal \cite{Nori} we calculated the exact reduced density $\rho_{S}(t)$ via the formula
\begin{equation}
\hat{\rho}_{S} (t)=\sum_{n=1}^{20} p_{n} {\rm Tr}_{B} 
[ |\Psi_{n}(t) \rangle \langle \Psi_{n}(t)| ],
\label{TDen2}
\end{equation}
\noindent
where the populations of bath are defined by
\begin{equation}
p_{n}=\frac{e^{-E_{n}/kT}}{ \sum_{m=1}^{20} e^{-E_{m}/kT}}
\end{equation}
\noindent
and $|\Psi_{n}(t) \rangle$ are the solutions to the Schr\"{o}dinger equation for all initial states $|\Psi_n(0)\rangle=|\psi(0)\rangle\otimes | n\rangle$. The initial system state was chosen as a superposition state $|\psi(0)\rangle=(|0\rangle+|1\rangle)/\sqrt{2}$. $20$ bath eigenstates sufficed for all our calculations at the given low temperature. We performed numerical integrations by an eighth order variable stepsize Runge-Kutta method\cite{RK} and exact diagonalization of the bath Hamiltonian is achieved by a Lanczos algorithm\cite{Arp} for $N=10$ bath qubits.  
  
\subsection{Chaotic Kraus decomposition approach}

The explicit form of Kraus operators for the Hamiltonian of our model is 
\begin{equation}
\hat{\cal K}_{n}(t)=\sqrt{p_{n}}e^{\frac{i}{\hbar}(\frac{1}{2}B_{z}^{(0)}\hat{\sigma}_{z}-B_{n,n}\hat{\sigma}_{x}^{(0)})t}.
\end{equation}
The chaotic Kraus decomposition for the given temperature $kT$ and initial density of the form $\hat{\rho}(0)=|\psi(0)\rangle \langle \psi (0)|$ then has the form
\begin{eqnarray}
\hat{\rho}_{S}(t) = \sum_{n=1}^{20} p_n \left(\begin{array}{cc} |c_{1}^{n}(t)|^{2} &  c_{1}^{n}(t)[c_{0}^{n}(t)]^{\ast} \\ c_{0}^{n}(t)[c_{1}^{n}(t)]^{\ast} & |c_{0}^{n}(t)|^{2} \end{array}\right)
\end{eqnarray}  
where 
\begin{eqnarray}
c_{1}^{n}(t) = \frac{\sqrt{2}}{2} \left[ \cos{a_{n} t} + i \frac{(B_{z}-2B_{n,n})}{b_{n}} \sin{a_{n} t}  \right] \\
c_{0}^{n}(t) = \frac{\sqrt{2}}{2} \left[ \cos{a_{n} t} - i \frac{(B_{z}+2B_{n,n})}{b_{n}} \sin{a_{n} t}  \right]
\end{eqnarray}

\begin{equation}
~a_{n}=\frac{b_{n}}{2\hbar}~~~~ {\rm and}~~~~ b_{n}=\sqrt{B_{z}^{2}+4 B_{n,n}^{2}}
\end{equation}

We calculated the diagonal matrix elements of the bath coupling operator by using the exact eigenstates of the bath Hamiltonian.

\subsection{Test for chaos}

\begin{figure}[t]
\centering
\includegraphics[scale=0.27]{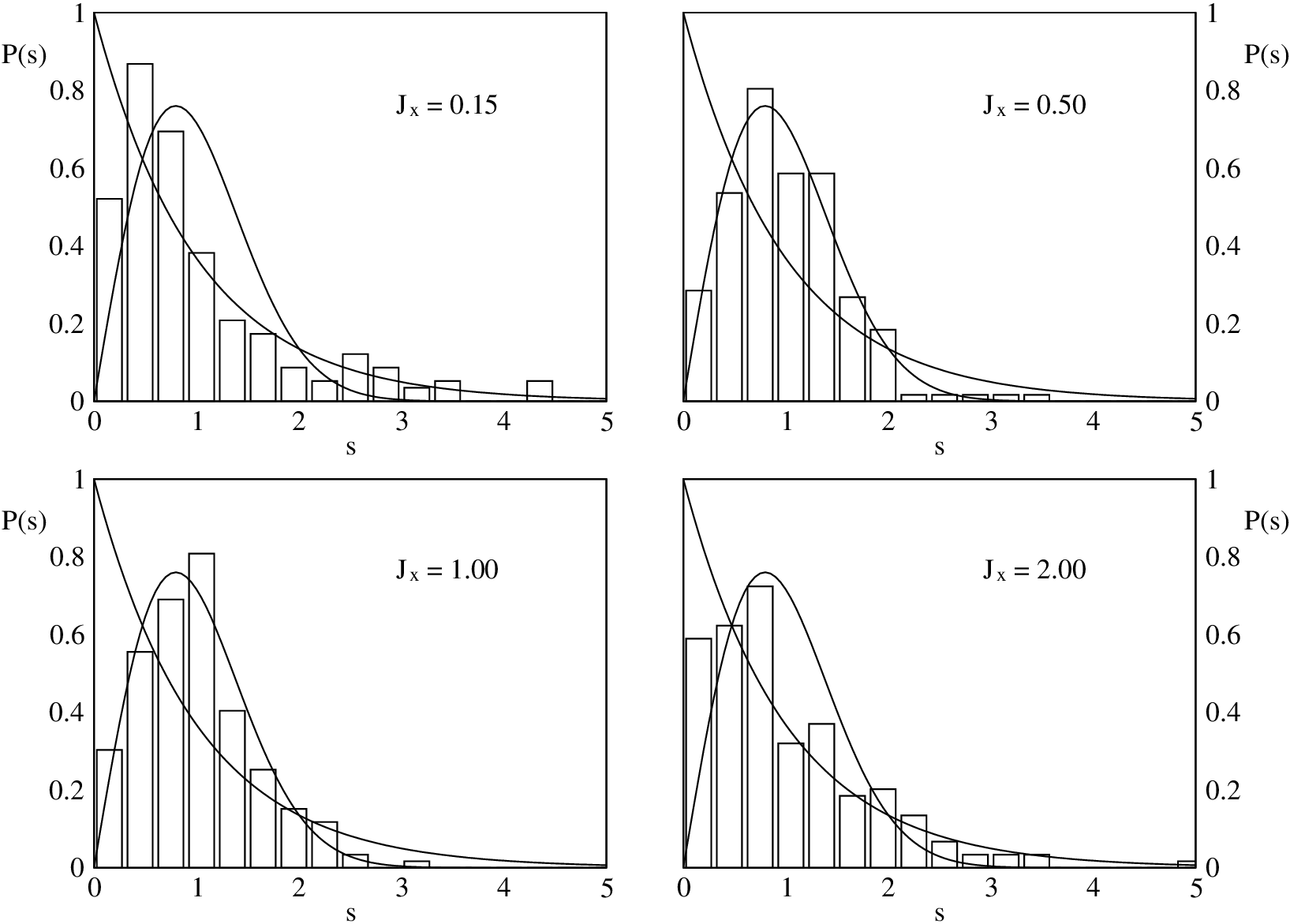}
\caption{Level spacing statistics for chaotic bath Hamiltonian}
\label{level}
\end{figure}

In order to confirm an integrable to chaotic transition in the bath Hamiltonian $\hat{H}_{B}$ as a function of $J_{x}$, we analyzed the nearest neighbor level spacing distribution\cite{Chaos} $P(s)$ for the 200 unfolded lowest eigenvalues. The unfolded spectrum  $\bar{E}_i$ is obtained from the actual eigenspectrum $E_i$ via smoothed staircase functions $\bar{N}(E)$, i.e. $\bar{E}_i=\bar{N}(E_i)$. Smoothed staircase functions are polynomial best fits to the actual staircase functions $N(E)=\sum_{i=1}^{200}\Theta(E-E_i)$ where $\Theta(x)$ is a Heaviside step function. The onset of chaos is indicated by the functional form of $P(s)$ changing from the Poisson form $\exp{(-s)}$ characteristic of 
 integrable systems to the Wigner--Dyson form $(\pi/2)s \exp{(-\pi s^{2}/4)}$ associated with chaos. Our results are given in Figure \ref{level}. The onset of chaos is observed above $J_{x}=0.15\ \epsilon$. For stronger couplings the eigenstatistics are consistent with Wigner--Dyson distribution. 

While this level spacing statistic is considered to be a universal indicator of quantum chaos, it does not provide information on the degree of chaoticity. We wish to know how the accuracy of the Kraus decomposition changes with increasing degree of chaos. Therefore, we alternatively refer to the Loschmidt echo $M(t)$\cite{Peres}, which is also widely believed to be an efficient indicator of quantum chaos\cite{Emerson}. 

\begin{figure}[t]
\centering
\includegraphics[scale=0.3]{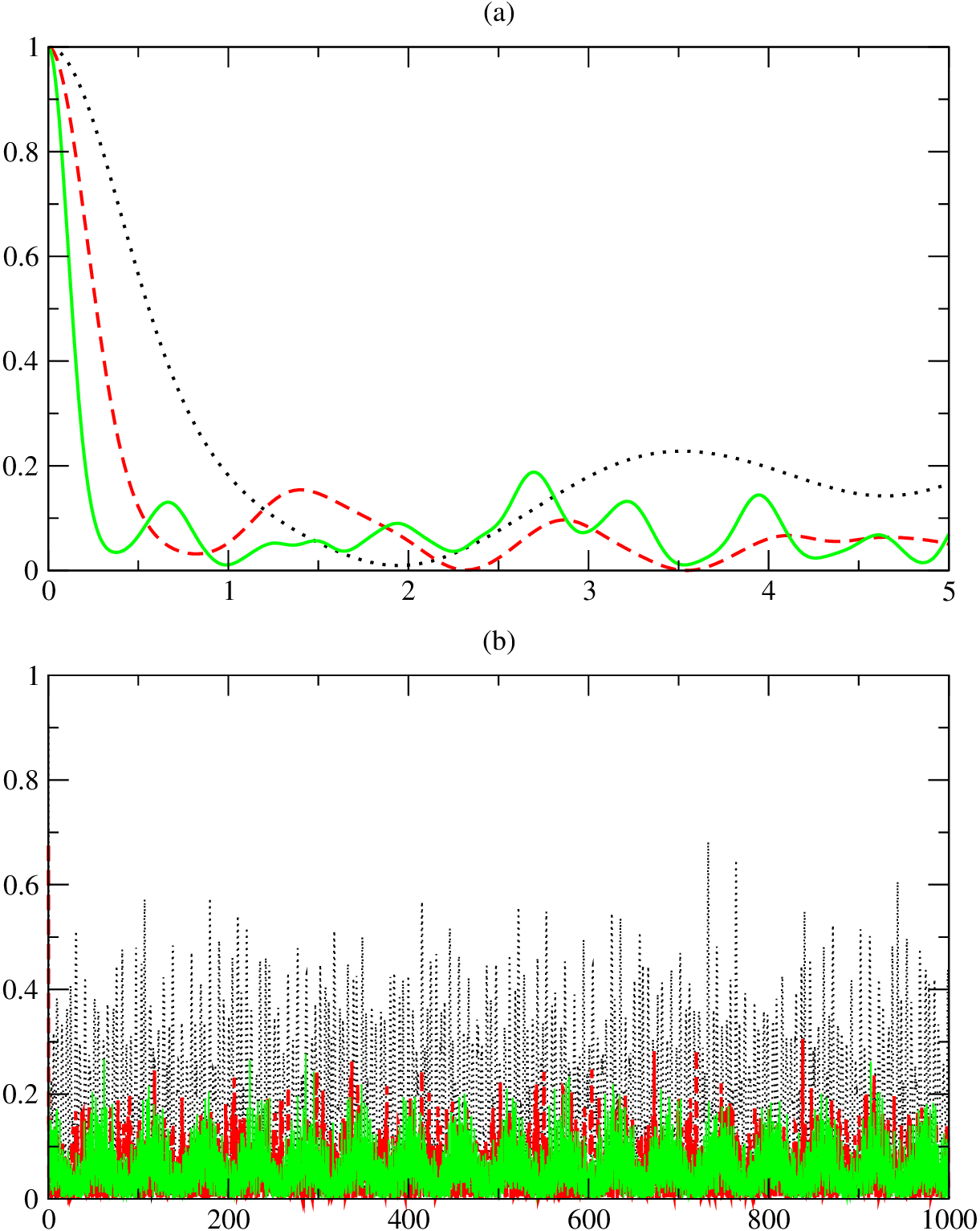}
\caption{(a) Short and (b) long time Loschmidt echo dynamics of bath Hamiltonian for $J_{x}=0.50$ (black solid line), $J_{x}=1.00$ (red dotted line), and $J_{x}=2.00$ (green dashed line).}
\label{echo}
\end{figure}

We calculate $M(t)$ for the bath Hamiltonian with the following formula
\begin{equation}     
M(t)=| \langle \psi_{0}| \exp{\{ i\hat{H}_{0}t/\hbar \}}\exp{\{-i(\hat{H}_{0}+\hat{V})t/\hbar\}}|\psi_{0}\rangle|^{2}
\end{equation}
where $|\psi_{0}\rangle$ is an initial state chosen as the ground state of $\hat{H}_0$. $\hat{H}_{0}$ is the integrable bath Hamiltonian (i.e. $\hat{H}_B$ for $J_{x}=0.00$) and $\hat{V}$ is the chaos generating perturbation Hamiltonian for $J_{x}=0.50,1.00,2.00$. A summary of $M(t)$ calculations is presented in Fig. (\ref{echo}).  
It is clear from Fig. (\ref{echo}) that increasing the magnitude of intra-bath coupling $J_{x}$ results in faster exponential decay of $M(t)$, which can be interpreted as increasing the degree of chaoticity.   

\subsection{Quantitative measures of environmental effects}

We use purity, ${\cal{P}}(t)={\rm Tr}_{S}[ \hat{\rho}_{S}(t) ]^{2}$, to determine the extent of non-unitary evolution in the system (i.e. decoherence and dissipation). The ideal value of purity is unity for pure initial system states. The fidelity is defined by ${\cal{F}}(t)={\rm Tr}_{S}[ \hat{\rho}_{S}(t)\hat{\rho}_{S}^{ideal}(t) ]$ where $\hat{\rho}_{S}^{ideal}(t)$ is ideal system evolution in the absence of system-bath interactions. The ideal value of fidelity is also 1. While the purity is insensitive to the unitary effects of system-bath interactions, the fidelity is sensitive to both unitary and non-unitary effects. Therefore, large deviations between the magnitudes of purity and fidelity are a good indicator of the presence of unitary 
shifts.

\section{Results} 
We tested the accuracy of the chaotic Kraus OSR for several values of intra-bath couplings in the chaotic bath regime. Here we compare the prediction of the chaotic Kraus decomposition with exact numerical results by calculating the purity ${\cal P}(t)$ and fidelity ${\cal F}(t)$ of the reduced density.

\begin{figure}
\centering
\includegraphics[width=3in,height=3.5in]{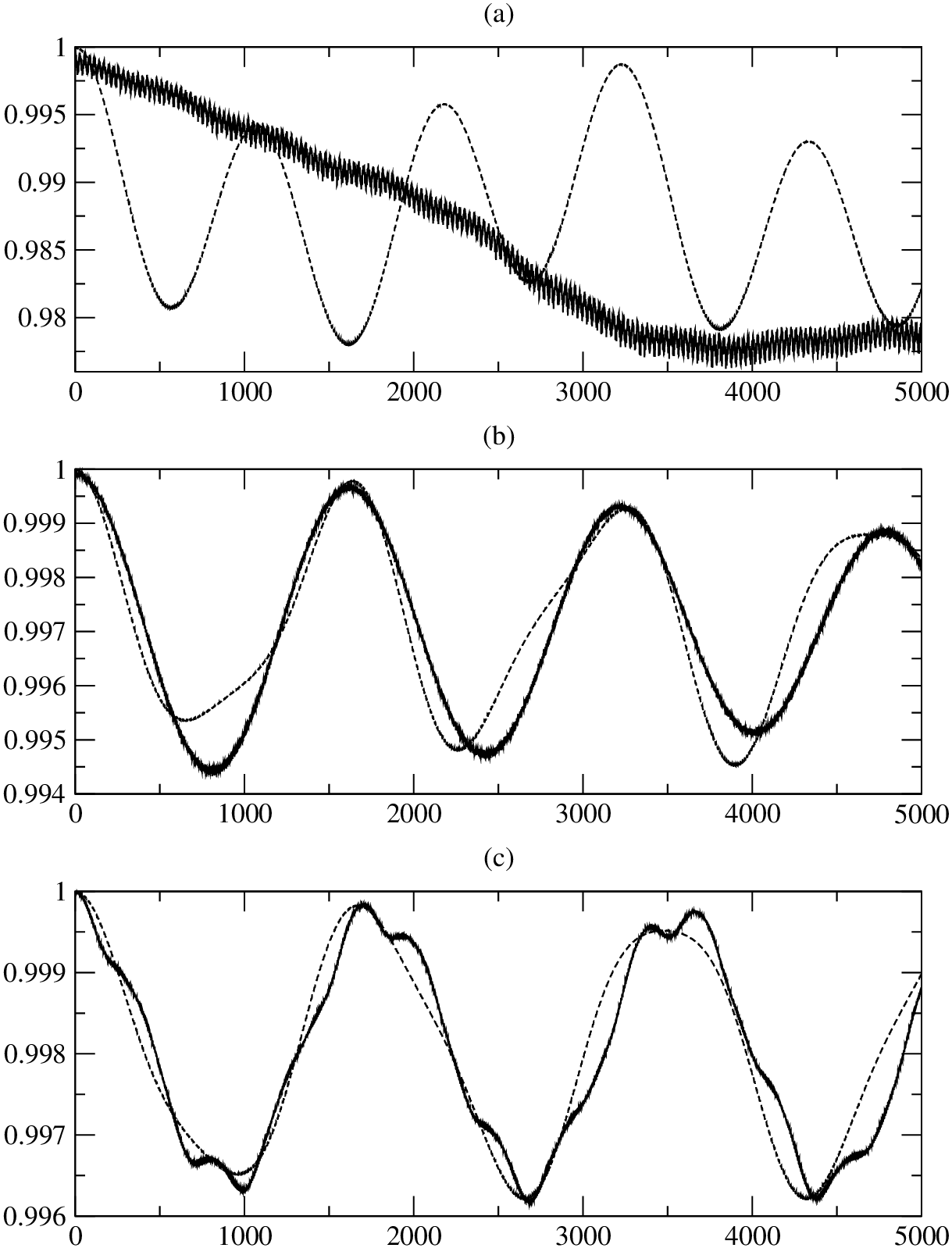}
\caption{Exact numerical (solid lines) and Kraus (dotted lines) results for purity ${\cal P}(t)$: for $J_{x}=0.50$ in (a), $J_{x}=1.00$ in (b), $J_{x}=2.00$ in (c).}
\label{pur}
\end{figure}

In Figure (\ref{pur}) we plot ${\cal P}(t)$ for three different values of intra-bath coupling $J_{x}$. (Time is in units of $\hbar/\epsilon$.)
The dashed lines are the Kraus predictions while the solid curve represents the exact dynamics. Even for $J_{x}=0.5\ \epsilon$, shown in part (a), the decoherence predicted by the Kraus decomposition is of the correct order of magnitude. Clearly, however it is not quantitatively accurate. There is also a faster time scale to the exact dynamics which is not captured at all by the Kraus decomposition. In part (b), for $J_{x}=1\ \epsilon$ and hence stronger chaos, we see much better agreement. In part (c), for $J_{x}=2\ \epsilon$, the Kraus decomposition is in very good agreement with the exact results. This is remarkable given that the bath contains only ten qubits. Moreover, it suggests that the 
Kraus decomposition could actually be employed as a useful computational method for low temperature systems.

\begin{figure}
\centering
\includegraphics[width=3in,height=3.5in]{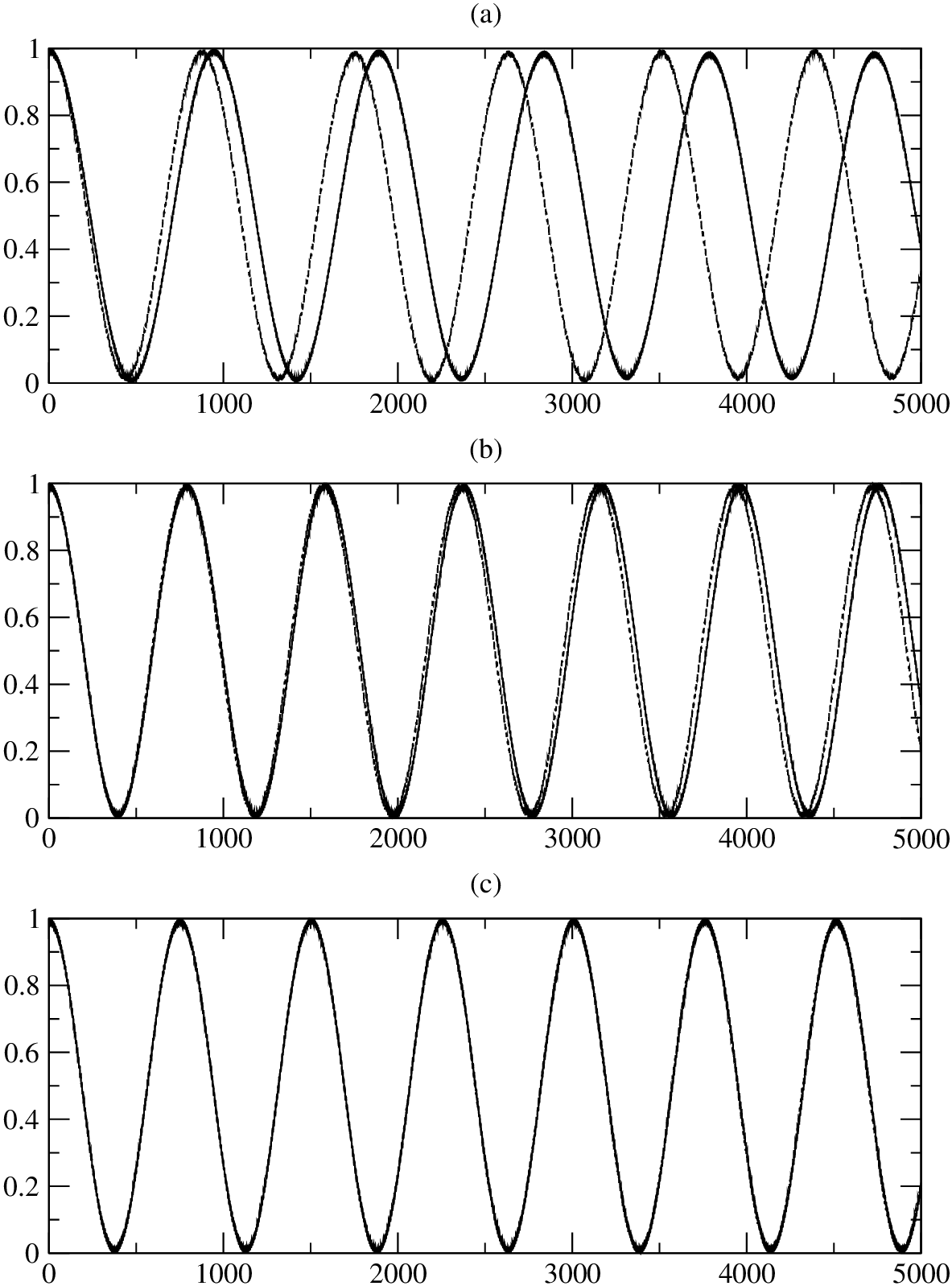}
\caption{Exact numerical (solid lines) and Kraus (dotted lines) results for fidelity ${\cal F}(t)$: for $J_{x}=0.50$ in (a), $J_{x}=1.00$ in (b), $J_{x}=2.00$ in (c).}
\label{fidel}
\end{figure}

In Figure (\ref{fidel}) we plot ${\cal{F}}(t)$ for both exact and Kraus decomposition results for the same three different values of intra-bath coupling $J_{x}$. Notice that the magnitude of the errors in ${\cal{F}}(t)$ is much larger than those in the purity ${\cal P}(t)$. The purity measures only non-unitary errors. ${\cal F}(t)$ is also sensitive to unitary errors. The large deviation of ${\cal P}(t)$ from unity therefore
indicates the presence of a large coherent shift. The origin of this shift has been explained in detail elsewhere\cite{CW2}. The agreement 
between the Kraus and exact results is quite good even for $J_{x}=0.5\ \epsilon$ (shown in part (a)). For $J_{x}=1\ \epsilon$ (part (b)) and
$J_{x}=2\ \epsilon$ (part (c)), the agreement is excellent.

\section{Discussion}

The Kraus decomposition captures the correct quantitative behavior for both unitary and non-unitary types of contributions from chaotic environmental dynamics. The accuracy of the Kraus decomposition even for our small bath model is quite impressive and thus it it may prove a useful numerical simulation method. The suggestive form of the decomposition (\ref{CTK}) is the result of primary interest. For example, it suggests that all non-unitary
effects vanish when all diagonal coupling matrix elements are identical, i.e. $B_{k,k}=B$.  In this case the dynamics reduces to
\begin{equation} 
\hat{\rho}_{S} (t)=  e^{-\frac{i}{\hbar}(\hat{H}_{S}+\hat{S}B)t}\hat{\rho}_{S}(0) e^{\frac{i}{\hbar}(\hat{H}_{S}+\hat{S}B)t}
\end{equation}
and the only effect of the bath is the coherent shift $\hat{H}_{S} \rightarrow \hat{H}_{S}+\hat{S}B$. 

Moreover, it is also clear
that the greater the variance in the coupling matrix elements, the more decoherence will be observed. This prediction is in
agreement with that of an approximate master equation\cite{CW1,CW2,SRA1,SRA2,SRA3} which has been shown to be quantitatively accurate
for a number of systems\cite{CW1,CW2}. 

It also predicts that a chaotic bath of thermodynamic dimension with uncorrelated initial state cannot cause 
decoherence or dissipation at absolute zero temperature. Such a bath can still cause coherent shifting, however. It was previously known that strongly chaotic finite baths 
cause little decoherence at low temperature\cite{Tess,Chaobath}. The fact that no decoherence is possible at zero temperature for large chaotic baths
is in strong contrast to the behavior of integrable baths. In the spin-boson model strong decoherence is possible even
when the bath has zero temperature.

\section{Summary}

We derived an exact and explicit Kraus decomposition for systems interacting with chaotic baths of thermodynamic dimension. We showed that
the decomposition is very accurate even for a small finite qubit bath, and it predicts all important features of the open system dynamics induced by chaotic environments. It could therefore prove useful as a numerical simulation method.

\ack{The authors gratefully acknowledge the support of the Natural Sciences and Engineering Research Council of Canada and computing resources provided by WestGrid.}

\end{document}